\documentclass[twocolumn,superscriptaddress,floatfix,pra,showpacs]{revtex4}
\usepackage{amsmath,amssymb,bm}
\usepackage{graphicx}
\usepackage{float}
\usepackage{color,dsfont}
\usepackage{appendix}

\begin{document}

\title{Long-distance quantum communication \\ through any number of entanglement-swapping operations}

\author{Aeysha Khalique}
\affiliation{School of Natural Sciences, National University of Sciences and Technology,
	H-12 Islamabad, Pakistan}
\affiliation{Hefei National Laboratory for Physical Sciences at Microscale and Department of Modern Physics, University of Science and Technology of China, Hefei, Anhui 230026, China}
\affiliation{Shanghai Branch, CAS Center for Excellence and Synergetic Innovation Center in Quantum Information and Quantum Physics, University of Science and Technology of China, Shanghai 201315, China}
\author{Barry C. Sanders}
\affiliation{Hefei National Laboratory for Physical Sciences at Microscale and Department of Modern Physics, University of Science and Technology of China, Hefei, Anhui 230026, China}
\affiliation{Shanghai Branch, CAS Center for Excellence and Synergetic Innovation Center in Quantum Information and Quantum Physics, University of Science and Technology of China, Shanghai 201315, China}
\affiliation{Institute for Quantum Science and Technology, 
	University of Calgary, Alberta T2N 1N4, Canada}
\affiliation{Program in Quantum Information Science, Canadian Institute for Advanced Research, Toronto, Ontario M5G 1Z8, Canada}

\date{\today}

\begin{abstract}
We develop a theory and accompanying mathematical model for quantum communication via any number of intermediate entanglement swapping operations and solve numerically for up to three intermediate entanglement swapping operations.
Our model yields two-photon interference visibilities post-selected on photon counts at the intermediate entanglement-swapping stations. Realistic experimental conditions are accommodated through the parametric down-conversion rate, photon-counter efficiencies and dark-count rates, and instrument and transmission losses. We calculate achievable quantum communication distances such that two-photon interference visibility exceeds the Bell-inequality threshold.
\end{abstract}

\maketitle

\section{Introduction}
\label{sec:intro}
The quest for long-distance secure quantum-communication relies on 
the technology of quantum repeaters~\cite{DLCZ01}
or quantum relays~\cite{CGR05} en route.
Quantum-communication reach is scalable (qubit rate falls polynomially with respect to sender-receiver separation~$\ell$)
for repeaters and exponentially sacalable with respect to relays.
An alternative, relatively easy-to-implement quantum communication network uses trusted
repeaters~\cite{SPD+10,BS08} between the sender and receiver but is not a long-term solution
to security threats.

Quantum-repeater and -relay networks both rely on entanglement swapping~\cite{PBW+98} as a building block
by dividing the separation into segments and then establishing entanglement swapping units
in each of these segments.
For scalable quantum communication,
optical quantum memory~\cite{LST09}
is needed;
until then the quantum-relay provides a practical approach to secure long-distance communication.
Combined with satellite-based quantum communication~~\cite{AJP+03,VJT+08},
just a few entanglement swapping steps may be needed for secure continent-scale quantum communication.

Although the experimental state-of-the-art
is only a single entanglement swapping station~\cite{MRT+04,dRMvH+05}
up to 143~km~\cite{HSF+14},
the theory of practical entanglement-swapping-based quantum communication
accounting for sources with higher-order multi-photon events and inefficient detectors 
with dark counts~\cite{SHST09} has been developed
for~$N$ entanglement swapping operations with 
$N$ any power of~$2$~\cite{KTS13}.
As is typical for simulating general quantum systems,
the hugeness of Hilbert space mitigates against precise numerical solutions so this
model has only been tractable up to $N=2$ entanglement swapping segments~\cite{KTS13}.

Here we advance the theory of entanglement-swapping-based quantum communication by 
making the theory applicable to any number of entanglement-swapping stations instead of only for 
powers of 2.
Increasing the number of entanglement swapping stations by just one is a hero\"{i}c feat both 
experimentally and theoretically, with the latter being challenging because the number of modes and hence the size of the Hilbert space rises rapidly with the number of entanglement swapping stations.
The Hilbert-space dimension is~$(n_\text{max}+1)^{8N}$
for~$n_\text{max}$ the photon-number truncation for each of~$8N$ modes.

Therefore, a theory that isrestricted to stations whose number is a power of two
is too restrictive especially because the case of~$N=3$ entanglement swapping stations
is excluded from such a theory.
Not only does our theory include the $N=3$ case,
but we have found shortcuts in the calculations to push the numerics from struggling to simulate
$N=3$ stations to being able to simulate $N=3$ stations.
Therefore, we solve the expected $4N$-photon-coincidence visibility as function of separation
and source and detector parameters.
We also solve the $N=1$ and $N=2$ visiblities to compare with previous results~\cite{SHST09,KTS13},
and we demonstrate agreement.
At present $N=4$ is out of reach numerically,
and drastically new approaches are needed to surmount the $N>3$ simulation barrier.

This paper is organized as follows.
In Sec.~\ref{sec:resources} we give briefly review how to model the resources employed in our set up, namely sources, detectors and channels.
We also provide numerical values for parameters used in our numerical simulations.
In Sec.~\ref{sec:concatenating}, we develop the theory for any number of elementary 
entanglement swapping operations
between the two separated parties.
We present our closed-form solution for the resultant state at the outermost modes
held by the two distant parties.
This closed-form solution is given as a nested sum,
which can be used in principle to determine the 
$4N$-photon-coincidence probability.

In Sec.~\ref{sec:numerical},
we solve the photon-coincidence visibility for the $N=3$ case numerically
and provide details of numerical shortcuts and the computational method that enabled these calculations in reasonable computation time. We compare these visibilities with  
the previous results for~$N=1$ and for $N=2$ in Sec.~\ref{sec:discussion}.
We also compare the dependence of photon-coincidence visibilities 
as a function of source brightness and separations~$\ell$ for the $N=1$, $N=2$, and $N=3$ cases.
The communication distances are also compared for the three cases of concatenations.
We conclude in Sec.~\ref{sec:conclusions}.

\section{Resources}
\label{sec:resources}

We develop a theory for determining the state at the two end nodes of a quantum-relay network,
with the intermediate channel comprising a linear chain of~$N$ entanglement-swapping stations.
Our model includes the production of unwanted multiple pairs of photons,
detector inefficiencies and dark counts, 
and channel losses.
In this section we review the mathematical descriptions used to model these resources.

The mathematical description of parametric down conversion (PDC) sources is given in Sec~\ref{subsec:sources}. 
Our brief review of the detector model,
which accounts for non-unit efficiency and dark counts,
is given in Sec.~\ref{subsec:detectors}.
We explain how transmission losses and other constant losses are accommodated in the detector efficiency in Sec~\ref{subsec:transmission}.
We assign numerical values to these parameters in Sec.~\ref{subsec:parameters}
according to the current experimental conditions.
These parameters are used in our numerical simulations.

\subsection{Sources}
\label{subsec:sources}
At each entanglement station,
entangled photons are produced by a PDC.
We treat the entangled-photon source as a pure-state superposition of photon numbers
in four spatial modes ($a$, $b$, $c$, and~$d$) and two polarizations (H and V)
0 the vacuum state~$|\text{vac}\rangle$
by the transformation
\begin{equation}
\label{eq:2PDC}
	\left|\chi\right\rangle
		=\exp\left[i\chi(\hat{a}^\dagger_\text{H}\hat{b}^\dagger_\text{H}
			+\hat{a}^\dagger_\text{V}\hat{b}^\dagger_\text{V}
                        +\hat{c}^\dagger_\text{H}\hat{d}^\dagger_\text{H}
                           +\hat{c}^\dagger_\text{V}\hat{d}^\dagger_\text{V}+\text{H.c.})\right] 
                       \left|\text{vac}\right\rangle		
\end{equation}
with~$\chi$ proportional to the PDC nonlinear-optical $\chi^{(2)}$ value.
The photon-pair production rate is~$\chi^2$.
Higher-order terms such as photon four-tuples arise from a power-series expansion
of the exponential~(\ref{eq:2PDC}).

Expression~(\ref{eq:2PDC}) neglects imperfect pumping,
losses within the nonlinear source and polarization walk-off and chromatic dispersion.
However, losses within the nonlinear source can be taken into account as constant loss to be included in detector efficiency as discussed in Sec~\ref{subsec:detectors}.
Polarization walk-off and chromatic dispersion are increasingly more pronounced at long distances,
which we do model here but we leave this topic for future studies.

\subsection{Detectors}
\label{subsec:detectors}
Ideal detectors would count the photon number in each mode for a given polarization.
For example, consider the $H$-polarized $a$ mode.
The number operator is~$\hat{a}^\dagger_\text{H}\hat{a}_\text{H}$
with eigenstates~$\{|n\rangle_{a_\text{H}}\}$.
The detector would effect an operation that is mathematically equivalent to the 
projective operator $|n\rangle_{a_\text{H}}\langle n|$.

Although some detectors aim to count up to a few photons,
the usual detectors are of the threshold type,
which means that they cannot discriminate one from more than one photon~\cite{BS02}.
Thus, the ideal threshold detector has a binary projective measurement operator:
$\Pi=\{|\text{vac}\rangle\langle\text{vac}|,\mathds{1}-|\text{vac}\rangle\langle\text{vac}|\}$.
The spectrum of this projective-valued measure is $0$ for vacuum and $1$ for one or more photons.

\begin{figure}
	\includegraphics[width=\linewidth]{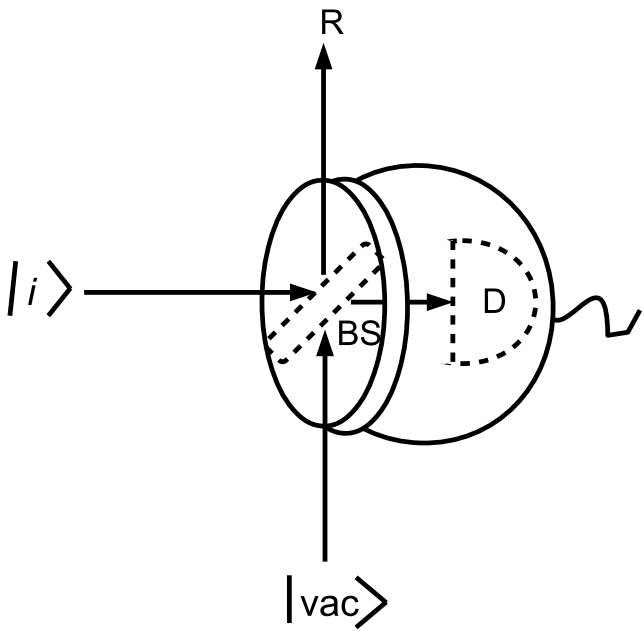}
\caption{
	An inefficient detector with efficiency $\eta_0$ is modeled as an efficient detector (D)
	preceded by a beam splitter (BS) with transmissivity $\eta_0$.
	The incident signal photons ($|i\rangle$) 
	combine with $|\text{vac}\rangle$ at the BS,
	and~$R$ is the reflected signal, which is traced out. 
     }
\label{fig:detector}
\end{figure}

In reality detectors are not perfectly efficient.
Sometimes a photon is missed and thus recorded as a zero,
i.e., as no count.
Detectors with intrinsic non-unit efficiency $\eta_0$ are modeled by a unit efficiency detector preceded by a fictitious beam splitter with transmissivity $\eta_0$~\cite{RR06} as shown in Fig.~\ref{fig:detector}.

The incident signal photons denoted by Fock state~$|i\rangle$
combine with the vacuum state $|\text{vac}\rangle$ at the beam splitter,
and the detected signal photons are obtained after tracing out the reflected part.
The conditional probability that~$q$ photons are detected given incident photons number~$i$ is
\begin{equation}
\label{eq:pqimodel}
	p(q|i)=\text{Tr}\left\{\Pi\text{Tr}_R\left[U_\text{BS}
		|i\rangle\langle i|\otimes|\text{vac}\rangle\langle \text{vac}|U_{\text{BS}}^\dagger\right]\Pi\right\},
\end{equation}
with~$U_\text{BS}$ being the unitary operator for the fictitious beam splitter.
Here $q=0$ represents no detection of a photon, and $q=1$ represents detection of one or more photons.

In addition to occasional missed detections of an incident photon,
a detector can click when there is no signal photon incident on it,
which is a dark count.
These clicks can be due to stray photons in the environment, so they are modeled by replacing $|\text{vac}\rangle$ in Fig.~\ref{fig:detector} and Eq.~(\ref{eq:pqimodel}) by a thermal state. For the threshold detector with dark-count probability $\wp$, such a detector model yields the conditional probability as \cite{SHST09}
\begin{align}
        p(q=0|i)
                =&(1-\wp)[1-\eta_0(1-\wp)]^i,
                          \nonumber\\
	p(q=1|i)
		=&1-p(q=0|i).
\label{eq:pqi}
\end{align}
The corresponding probabilities for photon number discriminating detectors are given in~\cite{SHST09}. For ideal threshold detectors, $\wp=0$ and $\eta=1$, which gives $p(q|i)=\delta_{q,i}$.

\subsection{Transmission losses}
\label{subsec:transmission}
Transmission losses and any other constant instrument loss 
encountered between source and detector, inclusively,
is incorporated into the detector efficiency parameter.
Transmission loss through a medium of length~$\ell$
with loss coefficient $\alpha$~dB/km is given as
\begin{equation} 
	\eta_t=10^{-\alpha\ell/{40N}}.
\end{equation}
The total length~$\ell$ is divided into segments of length $l/4N$ each,
which is the source-to-detector distance for each station.
The net efficiency of the detector is thus 
\begin{equation} 
	\eta=\eta_0\eta_t10^{-\alpha_0/10}
\label{eq:efficiency}
\end{equation}
with~$\alpha_0$ any other constant loss in the set-up.
The conditional probabilities in Eq.~(\ref{eq:pqi}) are now modified by replacing $\eta_0$ by $\eta$.

\subsection{Parameters}
\label{subsec:parameters}

For our numerical simulations, we employ certain values for the parameters discussed above.
The source brightness $\chi^2$ is treated as an independent tunable variable in many of our simulations.
In distance-dependent simulations of entanglement verification, we fix $\chi=0.1$
because lower values of $\chi$ would result in an overly long experiment run time to get the desired number of counts on the detector while higher values give quite a low value of entanglement measure at long distances.

In simulations with fixed distance~$\ell$,
the detector efficiency is $\eta=0.04$, which, for intrinsic efficiency $\eta_0=0.70$, represents a distance of around 200~km between the sender and receiver for a single intermediate station.
Superconducting-nanowires detectors have an efficiency around~$\eta_0=0.70$, which has been used for PDC sources \cite{JFY+13}, and are a plausible candidate for near-future quantum communication experiments.

With the inclusion of non-zero constant loss $\alpha_0$, achievable distances decrease.
We have taken $\alpha_0=4$~dB, where simulations have been carried out at various distances. The distance dependent loss coefficient is taken to be $\alpha=0.25$~dB/km, which reflects that used for optical fibers today~\cite{GRT+02}.
The dark-count probability $\wp$ is taken to be $10^{-5}$ throughout our simulations, which is the dark-count rate of present-day detectors \cite{dRMvH+05}.

\section{Concatenating entanglement swapping stations}
\label{sec:concatenating}

We now develop our theory for communication
through an arbitrary number of entanglement-swapping stations.
To this end,
we first review the optical entanglement swapping procedure in Sec.~\ref{subsec:entanglementswapping}. 
We explain how different stations are conjoined by entanglement swapping processing (Sec.~\ref{sec:conjoining}), which would interlink the two farthest ends.

We use visibility as a figure of merit for coincidences at the end nodes with the sender and receiver, which is explained in Sec.~\ref{subsec:multi-photon}.
 We then develop the mathematical model required for calculation of visibility in Sec.~\ref{subsec:Analytical results},
which requires calculations for coincidence probabilities for all stations and for the end nodes.
We obtain closed-form solutions of these coincidence probabilities for arbitrary concatenation of swappings.

\subsection{Entanglement swapping}
\label{subsec:entanglementswapping}

\begin{figure}
	\includegraphics[width=\linewidth]{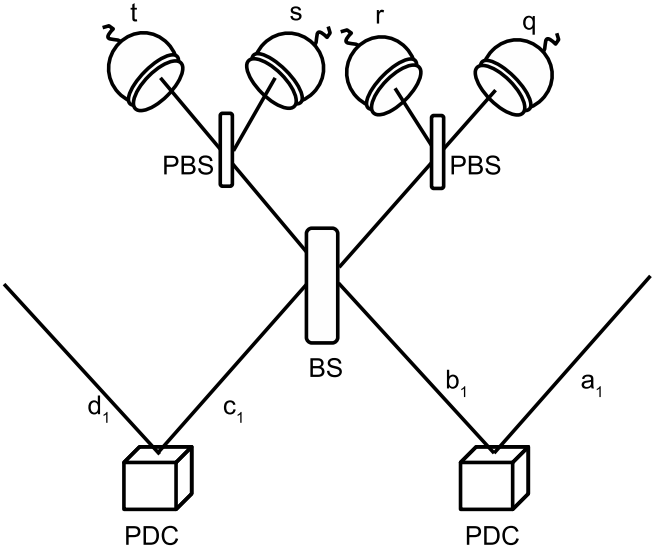}
\caption{
	Photons from adjacent PDC sources combine at a Bell measurement setup
	comprising a mode beam splitter (BS), a polarization beam splitter (PBS),
	and a four-tuple of detectors with clicks given by \{qrst\}.
     }
\label{fig:swap}
\end{figure}

The long-distance quantum relay network is divided into~$N$ entanglement-swapping stations.
In this way the total distance $\ell$ is partitioned into smaller segments each with a length~$\ell/N$.
Each station hosts an entanglement swapping set-up,
which is the same as that used in~\cite{SHST09}. 
The entanglement-swapping station comprises two PDC sources plus a Bell measurement set up made of a 50:50 beam splitter, polarization beam splitters and a four-tuple of detectors.
This entanglement-swapping station is shown in Fig.~\ref{fig:swap}. 

The two entangled states produced by the two PDC sources are given by Eq.~(\ref{eq:2PDC}).
Bell measurement is performed by combining the two innermost modes $b_1$ and $c_1$ in Fig.~\ref{fig:swap} on a beam splitter.
The four-tuple of detectors measure the photons in the two polarization modes of both $b_1$ and $c_1$.

In order to produce the singlet state $|\psi^-\rangle$, the four modes $\{b_H,b_V,c_V,c_H\}$ at the four-tuple of detectors must yield
either of the two four-tuple detection events~$\{1,0,1,0\}$ or~$\{0,1,0,1\}$.
These detection events ensure that the modes~$a$ and~$d$ are now entangled in the singlet state.
Thus entanglement is swapped according to $a\leftrightarrow d$ and $b\leftrightarrow c$ to $a\leftrightarrow d$ at each station. The modes $b$ and $c$ are measured and
hence no longer entangled. 

\subsection{Conjoining entanglement-swapping stations}
\label{sec:conjoining}

We now present our model for a relay set up and show how different stations are linked by entanglement swapping in a relay set up of~$N$ stations.

\begin{figure}
	\includegraphics[width=\linewidth]{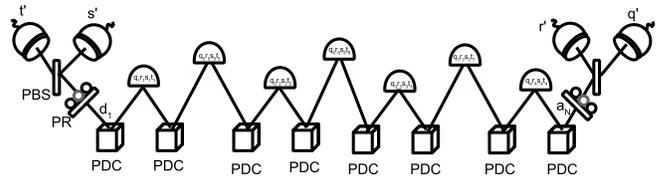}
\caption{
	Concatenation of~$N=4$ elementary swaps:
        Photons from adjacent PDC sources combine at a Bell measurement setup icluding four-tuple of detectrs with clicks given by $\{\bm{qrst}\}$, with $\bm{q}=(q_1q_2\dots q_{2N-1})$Each of the expressions for $\{\bm{r}$, $\{\bm{s}$ and $\{\bm{t}$, are also bit strings of length $2N-1$. This setup is represented by the caps. The two extreme left and right arm photons pass through polarization rotators (PR) and PBS before being detected by detectors.
     }
\label{fig:swaps}
\end{figure}

The entanglement swapping process explained in Sec.~\ref{subsec:entanglementswapping}, swaps the entanglement to the outermost modes of each station.
The adjacent stations are then linked together by performing a Bell measurement on the adjacent modes of the two stations as shown in Fig.~\ref{fig:swaps}.
Thus, for each $m\text{th}$ and adjacent $(m+1)\text{th}$ station, spatial modes $a_m$ and $d_{m+1}$ undergo Bell measurement, which swaps the entanglement to the leftmost mode, $d_m$,
of the $m\text{th}$ station, and the rightmost mode, $a_{m+1}$, of $(m+1)\text{th}$ station. 

Bell measurements are made by means of beam splitters, polarizer beam splitters and four-tuple of detectors as explained in Sec.~\ref{subsec:entanglementswapping} . The measurement events at the the $2N-1$ four-tuple of detectors are given by $\{\bm{q,r,s,t}\}$, with $\bm{q}=(q_1q_2\dots q_{2N-1})$. Each of the expressions for $\{\bm{r,s,t}\}$ are also bit strings of length $2N-1$. When all the adjacent stations are linked by Bell measurements resulting in the singlet state, entanglement is swapped to the extreme end modes, which are with the sender and the receiver. The whole entanglement swapping process requires $2N-1$ Bell measurements.
  
\subsection{Multi-photon coincidence visibility}
\label{subsec:multi-photon}

Having swapped the entanglement at the outermost modes of the relay set up we are now in a position to measure coincidences of photon counts at each station. These coincidences are measured as the conditioned probability that the pair of detectors at the end modes $d_1$ and $a_N$ both give a click,  given that the four-tuple of detectors at all the stations, yield the two photon click.
Thus, there are $4N$ photon coincidence counts required, namely two for each detector four-tuple.

Visibility is defined as the normalized difference of the maximum coincidence probabilities, $V_\text{max}$ and the minimum coincidence probabilities, $V_{\text{min}}$. Hence,
\begin{equation}
	V=\frac{V_\text{max}-V_\text{min}}{V_\text{max}+V_\text{min}}.
\label{eq:visibility}
\end{equation}
Here maximum counts are there when the outer two and all inner four-tuple of detectors yield the same singlet state i.e. record events~$\{1,0,1,0\}$ or~$\{0,1,0,1\}$.
The minimum count is taken to be the case where the outer two detectors record event~$\{0,1,1,0\}$ or~$\{1,0,0,1\}$.

Experimentally, this minimum count is achieved by introducing polarization rotators (PR) in the spatial modes $a$ and $d$, as shown in Fig.~\ref{fig:swaps}.
One of the rotators is kept at a fixed angle,
and $V_\text{max}$ and $V_\text{min}$ are calculated as a function of the angle of other rotator. When the two rotators are at the same angles then the values of maximum and minimum counts are recorded. 

For ideal detectors, with unit efficiency and no dark counts, irrespective of the number of stations, the state at spatial modes $a$ and $d$, after measuring {\color{red} the} singlet state at all stations, is given as
\begin{equation} 
	\left|\Phi_{a,d}\right\rangle
		=\frac{1}{\sqrt 2}\left(\frac{|1010\rangle-|0101\rangle}{\sqrt 2}+\frac{|0011\rangle-|1100\rangle}{\sqrt 2}\right),
\label{eq:perfect}
\end{equation}
where the first two terms will result in state $|\psi^-\rangle$ and will yield unit visibility. The other two terms correspond to a rejected event in coincidence probability. The multi-pairs from the source do not affect visibility for ideal detectors as each event in Eq.~(\ref{eq:perfect}) has identical dependence on~$\chi$.

\subsection{Coincidence probabilities for $N$ concatenated entanglement swapping stations}
\label{subsec:Analytical results}
Equipped with the model of our relay set up, we explain now how the coincidence probabilities are calculated mathematically.
At each detector the four-tuple count $\{q,r,s,t\}$ represents the observed count and $\{i,j,k,l\}$ represents the actual incident photons. The conditional probability to observe the event $\{q',r',s',t'\}$ on modes~$a_{N,H}$,~$a_{N,V}$, $d_{1,V}$ and~$d_{1,H}$ with non-ideal detectors, given Bell-state measurement events $\{\bm{q,r,s,t}\}$ at the $2N-1$ detector four-tuple, is
\begin{align}
	Q:=&p(q'r's't'|\bm{qrst})\nonumber\\
       =&\sum_{i',j',k',l'=0}^{\infty}p(q'r's't'|i'j'k'l')\nonumber\\
         &\times\prod_{u=1}^{2N-1}\sum_{i_u,j_u,k_u,l_u=0}^{\infty}|A^{\bm{ijkl}}_{i'j'k'l'}|^2\times    
          P^{\bm{qrst}}_{\bm{ijkl}}.
\label{eq:Q}
\end{align}
This equation is used to calculate the $2N$-photon coincidence probability. Here $|A^{\bm{ijkl}}_{i'j'k'l'}|^2$ is the transition probability of having actual incidences of~$\{i'j'k'l'\}$ on outer modes when there are actual incidences of~$\{\bm{ijkl}\}$ on the inner $(2N-1)$ four-tuple of detectors, wHere 
$\bm{i}=(i_1i_2\dots i_{2N-1})$, $\bm{j}=(j_1j_2\dots j_{2N-1})$, $\bm{k}=(k_1k_2\dots k_{2N-1})$ and~$\bm{l}=~(l_1l_2\dots \l_{2N-1})$.

$P^{\bm{qrst}}_{\bm{ijkl}}$ is the probability that ideal detectors would detect event $\{\bm{ijkl}\}$ given that an actual detection event yields the outcome $\{\bm{qrst}\}$. This probability is obtained by employing the Bayesian approach, which yields
\begin{equation}
	P^{\bm{qrst}}_{\bm{ijkl}}
		=\frac{p(\bm{qrst}|\bm{ijkl})p(\bm{ijkl})}
			{\prod\limits_{u=1}^{2N-1}\sum\limits_{i_u,j_u,k_u,l_u=0}^{\infty}
				p\left(\bm{qrst}|\bm{ijkl}\right)p(\bm{ijkl})}
\label{eq:P}
\end{equation}
With all detectors being independent $p(\bm{qrst}|\bm{ijkl})$ is the product of probabilities given by Eq.~(\ref{eq:pqi}). Here $P(\bm{ijkl})$ is the probability that ideally $\{ijkl\}$ photons are measured on inner modes after Bell measurement with detectors with unit efficiency and no dark counts. We refer to this as an ideal Bell measurement.

The quantum state at the extreme end modes ~$d_1$ and~$a_N$ after actual readout $\{\bm{q,r,s,t}\}$, at the $2N-1$ four-tuples of detectors at inner arms is
\begin{equation}
\rho=\prod_{u=1}^{2N-1}\sum_{i_u,j_u,k_u,l_u=0}^{\infty}P^{\bm{qrst}}_{\bm{ijkl}}\left|
      {\Phi_{\bm{ijkl}}}\right\rangle\left\langle{\Phi_{\bm{ijkl}}}\right|,
\label{eq:rho}
\end{equation}
where $P^{\bm{qrst}}_{\bm{ijkl}}$ is given in Eq.~(\ref{eq:P}).
The unnormalized state $\left|\Phi_{\bm{ijkl}}\right\rangle$ at the extreme modes after ideal readout $\bm {ijkl}$ at the inner detectors is given in Appendix~\ref{app:Phi}.

With the end-modes states,
conditioned on the ideal Bell measurement given by Eq.~(\ref{eq:Phi}),
we are now able to calculate the transition probability coefficient, $|A^{\bm{ijkl}}_{i'j'k'l'}|^2$, of ideally detecting $\{i'j'k'l'\}$ photons at the end modes after they pass through the polarizer rotators. The explicit expression for transition probability coefficient $A^{\bm{ijkl}}_{i'j'k'l'}$ is given in Appendix~\ref{app:TransitionA}.

We now have the general expression for coincidence probability for any arbitrary~$N$ number of concatenated swapping.  In deriving this expression, swapping is performed at all adjacent stations simultaneously whereas for~$N$ limited to powers of two, it is done by combining two adjacent stations at one time \cite{KTS13}. The expression for $N=2$ remains the same in both cases as there is one swapping for adjacent stations. For higher~$N$, however, there is a simplified expression for products of  $\Omega(\mu_n,\lambda_n,i_{N+n},l_{N+n})$.  Although the number of swaps remains the same, yet the simplicity of the expression gives a clear picture of the actual swapping process being applied. In addition, it allows the inclusion of all integer~$N$, which allows us to calculate visibilities up to $N=3$ numerically.

\section{Numerical solution for up to three concatenated swaps}
\label{sec:numerical}

We have developed an analytical theory for any number of concatenated swaps in Sec.~\ref{subsec:Analytical results}.
This leads us to calculate numerically the four-photon coincidence probability and hence visibility for up to $N=3$, in Sec.~\ref{subsec:P4N=3}. We have employed various numerical shortcuts to make these simulations possible. These shortcuts are discussed in detail in Sec.~\ref{subsec:numericalshortcuts}.

\subsection{Multi-photon coincidence probability for $N=3$}
\label{subsec:P4N=3}
With our analytical result for an arbitrary number of swaps, we are able to determine the visibility for the $N=3$ case numerically. For this case there are three elementary swaps and five Bell measurements. 
Thus, there are twelve-photon coincidences required for successful generation of entangled state at outermost ends.

For Bell-state measurement events $\{1,0,1,0\}$ or $\{0,1,0,1\}$ on all five detector four-tuples, 
which yields the singlet state,
the conditional probabilities of recording the incidences $\{1,0,1,0\},$ $\{0,1,0,1\},$ $\{0,1,1,0\}$ and $\{1,0,0,1\}$ are given as $Q_{1010}$, $Q_{0101}$,  $Q_{0110}$ and $Q_{1001}$
from Eq.~(\ref{eq:Q}). 
\begin{figure}
	\includegraphics[width=\linewidth]{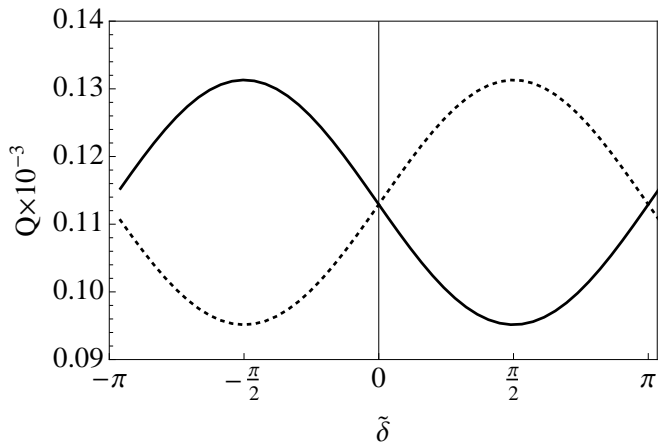}
\caption{
	The $4N$ coincidence probabilities vs polarization rotator angle $\tilde{\delta}$ for $N=3$: The coincidence probabilities $Q=Q_{1010}+Q_{0101}$ (dotted curve) are compared with $Q=Q_{0110}+Q_{1001}$ (solid curve) by varying $\tilde{\delta}$ and fixed angle $\tilde\alpha=\pi/2$. The detector efficiency is $\eta=0.04$ and dark-count probability is $\wp=1\times10^{-5}$. The truncation is done at maximum of three photons i.e.~$n_\text{max}=3$ per mode.
     }
\label{fig:MaxMinvsAngle}
\end{figure}
The maximum and the minimum coincidence rates~$Q_{1010}+Q_{0101}$ and~$Q_{0110}+Q_{1001}$, respectively, are plotted as a function of the polarization rotator angle $\tilde\delta$ for a fixed angle $\tilde\alpha=\pi/2$ in Fig.~\ref{fig:MaxMinvsAngle}.
The visibility is then calculated as maximum, $V_\text{max}$, and minimum, $V_\text{min}$, of these probabilities as given by Eq.~(\ref{eq:visibility})

We remark here that $\tilde\alpha$ and $\tilde\delta$ are the rotation angles of the Bloch vectors on the Bloch sphere. The rotation angles of the polarization vectors in the real space are half these angles and those of the $\lambda/2$ plates are one quarter of these angles.

As discussed in Sec.~\ref{subsec:parameters}, the dark-count probability is taken to be $\wp=1\times10^{-5}$ and detector efficiency is $\eta=0.04$. The source brightness is taken as $\chi^2=0.06$. 
This simulation reflects the experimental measurement process for visibility. As the reflection of characteristics of the singlet state, the coincidence probabilities for the anti-correlated polarizations $Q_{1010}+Q_{0101}$ is maximum when the two polarizer rotator angles are equal i.e. $\tilde\delta=\tilde\alpha=\pi/2$ and is minimum for $\tilde\delta=\tilde\alpha-\pi/2$. Similarly, as anticipated, the correlated polarizations, $Q_{0110}+Q_{1001}$, are maximum for $\tilde\delta=\tilde\alpha-\pi/2$ and minimum for $\tilde\delta=\tilde\alpha$. 
 
For the Fig.~\ref{fig:MaxMinvsAngle} simulations, visibility happens to be around 16\%.
The corresponding values of visibility for the $N=1$ and $N=2$ cases, are 70\% and 32\% respectively. Thus the decrease in visibility is less from $N=2$ to $N=3$ as compared to that from $N=1$ to $N=2$. A further comparison for different values of $\chi$ is shown in Sec.~\ref{subsec:comparing}. 

\begin{figure}
	\includegraphics[width=\linewidth]{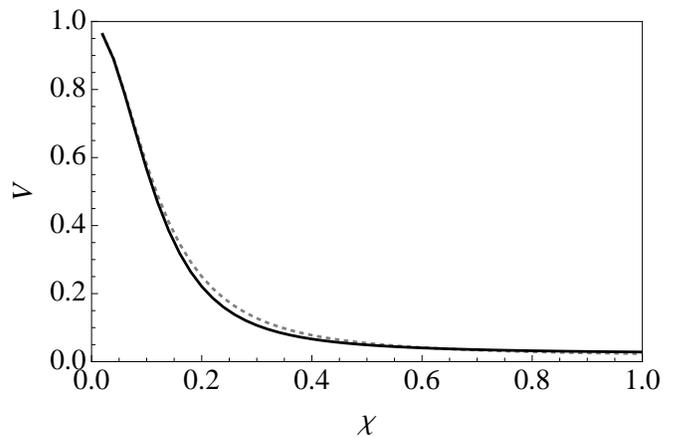}
\caption{
	Deviation in Visibility for~$n_\text{max}=2$ and~$n_\text{max}=3$:
	The visibility is shown for $N=3$ for truncation at~$n_\text{max}=2$ (dotted curve)
	and~$n_\text{max}=3$ (solid curve). 
     }
\label{fig:nmax2vs3n3}
\end{figure}

The truncation to lower values of~$n_\text{max}$ has the effect of higher visibility. However, the sinusoidal behavior of the maximum and minimum coincidence probabilities is present regardless of the choice of ~$n_\text{max}\ge1$. The truncation at~$n_\text{max}=3$ is reliable as shown in Fig.~\ref{fig:nmax2vs3n3}. There is little deviation from~$n_\text{max}=2$ to~$n_\text{max}=3$, hence for higher~$n_\text{max}$, the difference is even less.

\subsection{Numerical shortcuts}
\label{subsec:numericalshortcuts}

The dimension of our Hilbert space is $(n_\text{max}+1)^{8N}$, where~$n_\text{max}$ is the maximum number of photons in each mode. We have set~$n_\text{max}=3$, which makes the dimension of $N=3$ Hilbert space to be $10^5$ times more than that of $N=2$, and hence, the computation takes much longer than $N=2$. In order to obtain the results in reasonable computational time, we have applied various truncations on our numerical simulations, which we will discuss below. 

First we have kept~$n_\text{max}=3$ for our simulations. In the previous work, for   $N=2$~\cite{KTS13}, the truncation was done  at the same~$n_\text{max}$, however, for $N=1$~\cite{SHST09} simulations,
this maximum number could be raised to~$n_\text{max}=4$. As shown in Fig.~\ref{fig:nmax2vs3n3}, there is little deviation from~$n_\text{max}=2$ to~$n_\text{max}=3$ for $N=3$, and hence, it is a reliable truncation.

The transition probability $|A^{\bm{ijkl}}_{i'j'k'l'}|^2$ is small for a higher number of photons in each four-tuple of detectors, hence we have limited the sum of all the photons incident on each four-tuple of detectors to be not more than 4. This reduces the computational time significantly and was kept the same for $N=1$ and $N=2$ computation.

In addition, we have placed a lower bound on the sum of photons in each four-tuple of detectors. In order to have a coincidence click on two of the detectors in a detector four-tuple, there should be at least two photons arriving at the four detectors. For this purpose, we have excluded the events, where the sum of photons in each four-tuple of detectors is less than 2, as they will not lead to coincidence.

With all the above mentioned truncations, the dimension of the Hilbert space for $N=3$, with~$n_\text{max}=3$, has been reduced from around $10^{14}$ to $10^9$, as computed numerically.
Despite this reduction, 
an effective computer code code and appropriate computational techniques 
are needed to complete the computation in reasonable time as discussed in Appendix~\ref{app:computation}.

\section{Discussion}
\label{sec:discussion}

With the ability to numerically compute the conditional probability up to $N=3$, we can now compare variation of visibilities with the photon pair production rate for the three $N=1$, 2 and 3 cases. This comparison is done in Sec.~\ref{subsec:comparing}. We also compare the achievable distances for all three values of~$N$ in Sec.~\ref{subsec:VisibiltyvsDistance}.

\subsection{Comparing visibility}
\label{subsec:comparing}
\begin{figure}
	\includegraphics[width=\linewidth]{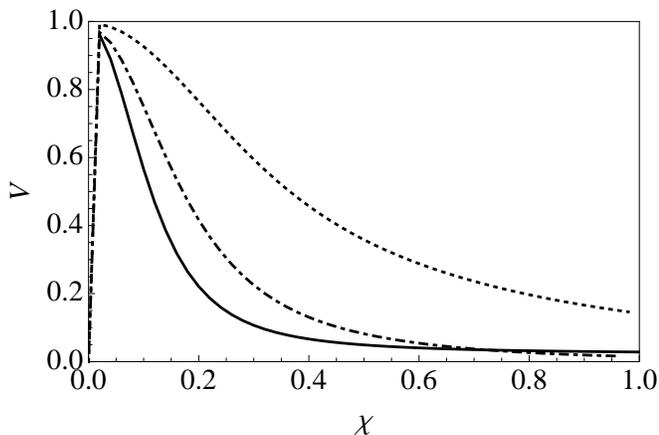}
\caption{
	Comparison of visibility for various number of elementary swaps:
        Visibility, $V$, is plotted vs the source efficiency $\chi$ for $N=1$ (dotted curve), $N=2$ (dot dashed curve) and $N=3$ (solid curve). The detector efficiency is $\eta=0.04$ and dark-count probability is $\wp=1\times10^{-5}$. The truncation imposes a maximum of three photons in each mode i.e.~$n_\text{max}=3$.
     }
\label{fig:N1N2N3}
\end{figure}

In Fig.~\ref{fig:N1N2N3}, we make the comparison of the variation in visibility with respect to the source efficiency $\chi$ for $N=1$, 2 and 3. The detector efficiency and dark-count rates are the same as in Fig.~\ref{fig:MaxMinvsAngle}.
Visibility decreases more rapidly as~$N$ is increased from $N=1$ to $N=3$, but the decrease is less from $N=2$ to $N=3$ as compared to that from $N=1$ to $N=2$.

For the former case the difference in visibilities is greater for lower values of~$\chi$. As shown earlier, the truncation at~$n_\text{max}=3$ is quite reliable.

\subsection{Achievable distances with concatenated swapping stations}
\label{subsec:VisibiltyvsDistance}

\begin{figure}
	\includegraphics[width=\linewidth]{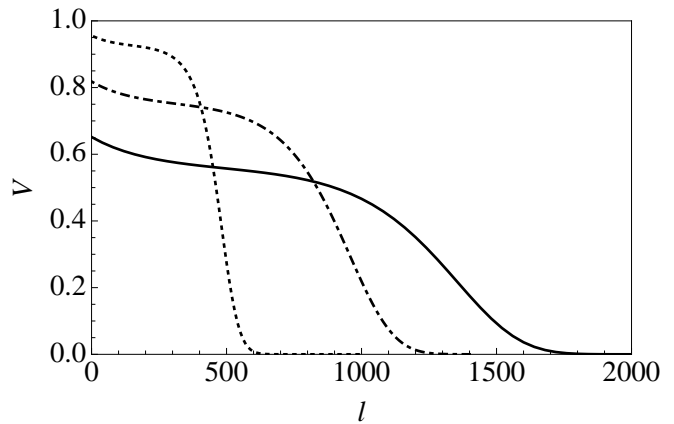}
\caption{
	Visibility vs communication distance: Visibility, $V$, is plotted vs distance, $\ell$, for $N=1$ (dotted curve), $N=2$ (dot dashed curve) and $N=3$ (black solid curve) for $\chi=0.1$.
     }
\label{fig:visibilityvsdistance}
\end{figure}

Quantum relays aim to deliver communication over long distances.
In order to assess the effect of increasing~$N$ on achievable distances for reasonable visibility,
we have compared them in Fig.~\ref{fig:visibilityvsdistance}.
The achievable distance is associated with transmission losses,
which are embedded in detector efficiency as given in Eq.~(\ref{eq:efficiency}).

In our simulations, various parameters are fixed as discussed in Sec.~\ref{subsec:parameters}. Thus the attained visibility is plotted for $N=1$, 2 and 3 vs distance (black curves) for $\chi=0.1$
in Fig.~\ref{fig:visibilityvsdistance}. 
Non-zero values of visibility are attained up to a distance of 600~km for $N =1$, 1200~km for $N=2$, and 1700~km for $N=3$. The visibility retains a significantly high value up to a certain distance and then steeply falls down as dark counts becomes effective.
 
The gain in achievable distance decreases with increasing~$N$.
For large~$N$, we expect that the achievable distance would saturate,
i.e., hit an upper bound. The effective communication distance bound saturates due to detector limitations, specifically inefficiency and dark counts, and to source limitations, specifically the random production of multiple pairs of photons in each event. We can see the effect of imperfect detection on bounding the effective communication distance by comparing the cases of ideal vs imperfect detectors. For ideal detectors, which have unit efficiency and zero dark-count rate, the computed visibility saturates close to unity over an asymptotically large distance. The slight deviation of visibility from unity in the ideal-detector case is due to source imperfections, namely multiple pairs of photons from either source. If every coincidence were due only to singlet states corresponding to a single entangled-photon pair from each source, ideal detection would yield unit-visibility coincidences. Unfortunately PDC sources are imperfect in that they deliver with low but non-negligible probability zero, one, two or more pairs per event, which leads to spurious coincidences that are insensitive to the polarizer rotator angle (abscissa in Fig.~\ref{fig:MaxMinvsAngle}), and hence diminish the maximum achievable visibility to less than unity.

Lower values of~$\chi$ are needed to have high visibility for a long distance.
However, reducing the source brightness in an actual experiment would increase the experiment run time to obtain reasonable number of coincidence counts for calculating visibility. Hence a  lower $\alpha_0$ and $\eta_0$ are needed to trade off between experiment run time and reasonable visibility to achieve larger distances.

\section{Conclusions}
\label{sec:conclusions}

We have developed a comprehensive theory for an arbitrary number of concatenations of entanglement swappings in a quantum relay setup. We have given a closed form solution of the state of the outermost modes for two distant observers conditioned on the Bell measurement of all inner concatenated modes including the imperfections.

Practical limitations include source brightness and detector dark counts and efficiency. The detector efficiency includes channel losses. We can hence calculate the $4N$ coincidence probability leading to successful Bell measurement.

We have applied various shortcuts to reduce the dimension of Hilbert space and hence the computation time. With these numerical shortcuts, we are able to calculate visibilities up to $N=3$. Our simulation provides a good upper bound on coincidence probabilities and hence visibilities.

We have compared the visibilities for $N=1$, 2 and 3 vs the source brightness.
As~$N$ increases, the visibility decreases rapidly.
Note that the rate of decrease in visibility diminishes with an increase in~$N$.

We have investigated the achievable distance, with non-zero visibility and for $\chi=0.1$.
This achievable distance increases  from 600~km for $N=1$ to 1200~km $N=2$ and to 1700~km for $N=3$. Thus the increase in achievable distance diminishes with increasing~$N$.
This is a significant point in that it suggests that,
beyond some limiting~$N$, an increase in distance cannot be achieved by increasing the number of concatenations. 

\acknowledgments
We acknowledge valuable discussions with P. Zhang and Q-C Sun,
and we appreciate financial support
from the 1000 Talent Program of China and from Alberta Innovates Technology Futures.
This research has been enabled by the use of computing resources
provided by WestGrid and Compute/Calcul Canada.

\appendix
\section{State $\left|\Phi_{\bm{ijkl}}\right\rangle$}
\label{app:Phi} 
The unnormalized state of the extreme end modes~$d_1$ and~$a_N$ of the relay after ideal readout $\bm{ijkl}$ at the $2N-1$ four tuple of detectors is
\begin{widetext}
\begin{align}
\left|\Phi_{\bm{ijkl}}\right\rangle
        =&\left(\prod_{p=1}^{N}\frac{1}
              {(\sqrt{2})^{i_p+j_p+k_p+l_p}\sqrt{i_p!j_p!k_p!l_p!}}\frac{(\tanh\chi)^{i_p+j_p+k_p+l_p}}
                {\cosh^{4N}\chi}
              \sum_{\mu_p=0}^{i_p}\sum_{\nu_p=0}^{j_p}\sum_{\kappa_p=0}^{k_p}
             \sum_{\lambda_p=0}^{l_p}(-1)^{\mu_p+\nu_p}{i_p\choose \mu_p}{j_p\choose 
                 \nu_p}
                 {k_p\choose \kappa_p}{l_p\choose \lambda_p}\right)\nonumber\\
      &\times\prod_{n=1}^{N-1}\Omega(\mu_n,
            \lambda_n,i_{N+n},l_{N+n})\Omega(\nu_n,
                \kappa_n,j_{N+n},k_{N+n})
           \frac{\sqrt{i_{N+n}!j_{N+n}!k_{N+n}!l_{N+n}!}}
            {(\sqrt{2})^{i_{N+n}+j_{N+n}+k_{N+n}+l_{N+n}}}\nonumber\\
        &    \times\delta_{i_{N+n}+l_{N+n},
           \mu_{n}+\lambda_{n}+i_{n+1}+l_{n+1}
              -\mu_{n+1}-\lambda_{n+1}}
             \delta_{j_{N+n}+k_{N+n},   
           \nu_{n}+\kappa_{n}+j_{n+1}+k_{n+1}
              -\nu_{n+1}-\kappa_{n+1}}\nonumber\\
     &\times\hat{d}_{1,\text{H}}^{\dagger i_1+l_1-\mu_1-\lambda_1}
                    \hat{d}_{1,\text{V}}^{\dagger j_1+k_1-\nu_1-\kappa_1}
                       \hat{a}_{N,\text{H}}^{\dagger \mu_N+\lambda_N}
                         \hat{a}_{N,\text{V}}^{\dagger \nu_N+\kappa_N}\left|
                          \text{vac}\right\rangle.
\label{eq:Phi}
\end{align}
\end{widetext}
Here
the $2(N-1)$ factors of~$\Omega$ in Eq.~(\ref{eq:Phi})
come from secondary Bell measurements, which connect the elementary adjacent swaps and are given as
\begin{align}
\label{eq:Omega}
	\Omega(\mu_n,
            \lambda_n,i_{N+n},l_{N+n})
		=&\sum_{\gamma=0}^{\mu_n+\lambda_n}{\mu_n+\lambda_n\choose \gamma}
			\nonumber\\&\times
		{{i_{N+n}+l_{N+n}-\mu_n-\lambda_n }\choose {i_{N+n}-\gamma}}
			\nonumber\\&\times
		(-1)^{\mu_n+\lambda_n-\gamma}.
\end{align}
The form of~$\Omega(\nu_n,\kappa_n,j_{N+n},k_{N+n})$ is analogous.

\section{Transition probability amplitude}
\label{app:TransitionA}
For polarizer rotators at angles $\tilde{\alpha}$ and $\tilde{\delta}$, for an arbitrary number of swappings,  the transition probability $A^{\bm{ijkl}}_{i'j'k'l'}$ is given as
\begin{widetext}
\begin{align}
A^{\bm{ijkl}}_{i'j'k'l'}=&
    \left(\prod_{p=1}^{N}\frac{1}{\sqrt{2^{i_p+j_p+k_p+l_p}i_p!j_p!k_p!l_p!}}\frac{(\tanh\chi)^{i_p+j_p+k_p+l_p}}       
         {\cosh^{4N}\chi}
        \sum_{\mu_p=0}^{i_p}\sum_{\nu_p=0}^{j_p}\sum_{\kappa_p=0}^{k_p}
         \sum_{\lambda_p=0}^{l_p}
            (-1)^{\mu_p+\nu_p}{i_p\choose \mu_p}{j_p\choose \nu_p}{k_p\choose \kappa_p}{l_p
                \choose \lambda_p}\right)\nonumber\\
    &\times\prod_{n=1}^{N-1}
        \Omega(\mu_{n},\lambda_{n},i_{N+n},l_{N+n})
            \Omega(\nu_{n}\kappa_{n},j_{N+n},k_{N+n})
       \frac{\sqrt{i_{N+n}!j_{N+n}!k_{N+n}!l_{N+n}!}}       
        {(\sqrt{2})^{i_{N+n}+j_{N+n}+k_{N+n}+l_{N+n}}}\nonumber\\
    &\times\delta_{i_{N+n}+l_{N+n},    
        \mu_{n}+\lambda_{n}+i_{n+1}+l_{n+1}-
         \mu_{n+1}-\lambda_{n+1}}
         \delta_{j_{N+n}+k_{N+n},
          \nu_{n}+\kappa_{n}+j_{n+1}+k_{n+1}-
            \nu_{n+1}-\kappa_{n+1}}\nonumber\\
    &\times(\nu_N+\kappa_N)!(j_1+k_1-\nu_1-\kappa_1)!\sqrt{\frac{j'!k'!}{i'!l'!}}
         \sum_{n_a=0}^{\text{Min}[j',\nu_N+\kappa_N]}
            \sum_{n_d=0}^{\text{Min}[k',j_1+k_1-\nu_1-\kappa_1]}
             \left(\text{i}\tan\frac{\tilde{\alpha}}{2}\right)^{\nu_N+\kappa_N+j'-2n_a}\nonumber\\
    &\times\left(\cos\frac{\tilde{\alpha}}
           {2}\right)^{i'+j'-2n_a}\left(i\tan\frac{\tilde{\delta}}{2}\right)^{k'+j_1+k_1-\nu_1-\kappa_1-2n_d}
              \left(\cos\frac{\tilde{\delta}}{2}\right)^{l'+k'-2n_d}
          \nonumber\\
   &\times\frac{(i'+j'-n_a)!(l'+k'-nd)!}{n_a!n_d!(j'-n_a)!(k'-n_d)!(\nu_N+\kappa_N-n_a)! 
           (j_1+k_1-\nu_1-\kappa_1-n_d)!}\nonumber\\
    &\times\delta_{i'+j',
         \mu_N+\nu_N+\kappa_N+\lambda_N}
\delta_{k'+l',i_1+j_1+k_1+l_1-\mu_1-\nu_1-\kappa_1-\lambda_1}.
\label{eq:A}
\end{align}
\end{widetext}

\section{Computational method}
\label{app:computation}
An efficient code was needed to minimize the computational time.
For this purpose, coding was done in c++.
To improve the efficiency of the code,
look-up tables have been constructed for various functions and products involved Eq.~(\ref{eq:A}) and hence~(\ref{eq:Q}). These included look-up tables for factorial function, combinations, and  $\Omega$ functions.

Care has been exercised to include products of powers of~$\sqrt{2}$ in the look-up tables because 
these calculations contribute substantially to computational time as the number of loops increases.
Also look-up tables have been made for the functions $p(q|i)$ in Eq~(\ref{eq:pqi}) and their products. The probabilities~$Q$ and $A_{i'j'k'l'}^{\bm{ijkl}}$ are called within the program as functions. 

For the $N=3$ case,
the code
which ran at 2.66 GHz on an Intel$^\circledR$ Xeon$^\circledR$ E5430 quad-core processor with 8~GB of memory,
required 9 hours to compute the $4N$-photon coincidence probability on a single core.
Hence, we parallelized our code and made use of multiple cores to calculate each coincidence probability  for various distances as well as source brightness.
This procedure required four processors to calculate visibility for a particular $\chi$ or $\ell$.


\end{document}